\documentclass[11pt,a4paper]{article}
\usepackage[margin=2.5cm]{geometry}
\usepackage{amsmath,amssymb,bm}
\usepackage{graphicx}
\usepackage{pdfpages}
\usepackage{booktabs}
\usepackage{multirow}
\usepackage[section]{placeins}  %

\usepackage[numbers,sort&compress]{natbib}
\usepackage{hyperref}
\hypersetup{colorlinks=true,linkcolor=blue!50!black,citecolor=blue!50!black,urlcolor=blue!50!black}
\usepackage{xcolor}
\newcommand{\todo}[1]{\textcolor{red}{[#1]}}
\newcommand{\alGapProdUniform}{0.17}
\newcommand{\alGapRfour}{0.27}
\newcommand{\alGapRone}{0.53}

\newcommand{\alRfourMAE}{0.40}

\newcommand{\alRiseRfour}{0.11}
\newcommand{\alRiseRone}{0.20}

\newcommand{\alRtwoMAE}{0.53}

\newcommand{\alRtwoTwistHigh}{1.11}
\newcommand{\alRtwoTwistLow}{0.37}

\newcommand{\bLEHTestTraj}{0.42}

\newcommand{\dimLEHAA}{0.17}
\newcommand{\dimLEHAB}{0.15}
\newcommand{\dimOMNI}{5.98}

\newcommand{\dimOMNIFT}{6.01}

\newcommand{\dimSM}{0.55}

\newcommand{\dimXM}{0.64}

\newcommand{\eLEHlarge}{3.9}

\newcommand{\eLEHrand}{5.1}

\newcommand{\eSMlarge}{3.7}

\newcommand{\eSMrand}{4.9}
\newcommand{\eXMlarge}{12.1}

\newcommand{\eXMrand}{4.0}

\newcommand{\fLEHrand}{0.21}

\newcommand{\fSMrand}{0.21}

\newcommand{\fXMrand}{0.27}
\newcommand{\fcLEHlarge}{0.13}
\newcommand{\fcLEHrand}{0.17}
\newcommand{\fcLEHrtwo}{0.92}

\newcommand{\fcXMlarge}{0.24}
\newcommand{\fcXMrand}{0.23}

\newcommand{\lcFoldMax}{49}
\newcommand{\lcFoldMin}{8}
\newcommand{\lcMolMax}{4}
\newcommand{\lcMolMin}{0.6}
\newcommand{\lgUniformLow}{0.73}
\newcommand{\lossAlllowgap}{44}
\newcommand{\lossAlltraj}{+0.3}
\newcommand{\lossAllval}{-3}
\newcommand{\lossLEHlowgap}{38}
\newcommand{\lossLEHtraj}{+25}

\newcommand{\monoLEH}{0.18}

\newcommand{\monoSM}{0.19}
\newcommand{\monoXM}{0.17}
\newcommand{\nALGeom}{68,119}
\newcommand{\nALMol}{16,381}

\newcommand{\nConjGeom}{12,770}
\newcommand{\nDimerPairs}{180}
\newcommand{\nDyes}{5,018}
\newcommand{\nDyesNoS}{3,875}

\newcommand{\nGeomAll}{303,061}
\newcommand{\nGeomExcAll}{85,950}
\newcommand{\nGeomFex}{203,040}
\newcommand{\nGeomTotal}{222,172}
\newcommand{\nLadders}{1,393}
\newcommand{\nLargeMol}{500}
\newcommand{\nMolAll}{106,933}

\newcommand{\nMolTotal}{100,522}

\newcommand{\nOutlier}{4}
\newcommand{\nOutlierTotal}{87,396}

\newcommand{\nRawCalc}{200,445}

\newcommand{\nTestScafMol}{10,003}
\newcommand{\nTrainExc}{12,869}
\newcommand{\nTrainGeom}{168,587}
\newcommand{\nTrainGeomProd}{228,594}
\newcommand{\nTrainMol}{76,621}
\newcommand{\nValMol}{1,500}

\newcommand{\oPairOther}{2.4}
\newcommand{\oPairOutMax}{32}
\newcommand{\oPairOutMin}{20}

\newcommand{\oProdLEHmin}{13}

\newcommand{\oProdSMmin}{21}
\newcommand{\pCommGainMax}{12}
\newcommand{\pCommGainMin}{2}
\newcommand{\pCorrMax}{0.99}
\newcommand{\pCorrMin}{0.97}

\newcommand{\pLEHTestTrajcomm}{0.29}

\newcommand{\pLEHValRise}{0.09}
\newcommand{\pLEHValTraj}{0.19}

\newcommand{\pLEHValTrajstd}{0.05}

\newcommand{\pLEHbiasMax}{0.54}
\newcommand{\pLEHbiasMin}{0.35}

\newcommand{\pLEHgapMax}{0.54}
\newcommand{\pLEHgapMin}{0.36}

\newcommand{\pLEHladSone}{0.43}
\newcommand{\pLEHladStwothree}{0.18}

\newcommand{\pLEHpotMax}{0.30}
\newcommand{\pLEHpotMin}{0.24}

\newcommand{\pLowGapLEHuni}{0.39}
\newcommand{\pLowGapLEHw}{0.25}
\newcommand{\pLowGapSMuni}{0.43}
\newcommand{\pLowGapSMw}{0.22}
\newcommand{\pRiseErrMax}{0.25}
\newcommand{\pRiseErrMin}{0.08}
\newcommand{\pRiseRefMax}{3.7}
\newcommand{\pRiseRefMin}{1.8}

\newcommand{\pSMTestTrajcomm}{0.15}

\newcommand{\pSMValRise}{0.07}
\newcommand{\pSMValTraj}{0.15}

\newcommand{\pSMValTrajstd}{0.01}

\newcommand{\pSMbiasMax}{0.22}
\newcommand{\pSMbiasMin}{0.11}

\newcommand{\pSMgapMax}{0.22}
\newcommand{\pSMgapMin}{0.18}

\newcommand{\pSMladSone}{0.32}
\newcommand{\pSMladStwothree}{0.30}

\newcommand{\pSMpotMax}{0.15}
\newcommand{\pSMpotMin}{0.08}

\newcommand{\pSmallTrajMax}{0.09}
\newcommand{\pStabDrift}{7}

\newcommand{\pctTrainLowGap}{2}
\newcommand{\pctTrainLowGapTwo}{7}

\newcommand{\popLadBase}{0.62}
\newcommand{\popLadPop}{0.62}
\newcommand{\popRandBase}{0.227}
\newcommand{\popRandPop}{0.215}

\newcommand{\popSvalBase}{0.67}

\newcommand{\popSvalPop}{0.78}

\newcommand{\qLEHrndten}{0.233}

\newcommand{\qLEHten}{0.256}
\newcommand{\qLEHtenall}{0.204}

\newcommand{\qLEHtenscaf}{0.291}
\newcommand{\qLEHval}{0.090}

\newcommand{\qOMNIten}{0.511}

\newcommand{\qOMNIval}{0.178}

\newcommand{\qRandNinePct}{83}

\newcommand{\qSMten}{0.265}
\newcommand{\qSMtenall}{0.229}

\newcommand{\qSMtenscaf}{0.285}
\newcommand{\qSMval}{0.092}

\newcommand{\qXMten}{0.849}

\newcommand{\qXMtenrtwo}{0.23}
\newcommand{\qXMval}{0.182}
\newcommand{\sepLEHfoursix}{0.24}
\newcommand{\sepLEHone}{0.13}
\newcommand{\sepLEHtwothree}{0.17}

\newcommand{\sepSMfoursix}{0.93}
\newcommand{\sepSMone}{0.20}
\newcommand{\sepSMtwothree}{0.73}
\newcommand{\sepSMunion}{0.15}

\newcommand{\tdshiftAAfive}{10}
\newcommand{\tdshiftAAshort}{65}

\newcommand{\tdshiftABfive}{4}
\newcommand{\tdshiftABshort}{31}

\newcommand{\uLEHTestTraj}{0.28}
\newcommand{\uSMTestTraj}{0.20}
\newcommand{\vALEHdyesRtwo}{0.17}
\newcommand{\vALEHdyesSone}{0.41}

\newcommand{\vALEHdyesStwothree}{0.24}

\newcommand{\vALEHladSone}{0.60}

\newcommand{\vALEHladStwothree}{0.24}
\newcommand{\vASMdyesRtwo}{0.04}
\newcommand{\vASMdyesSone}{0.39}

\newcommand{\vASMdyesStwothree}{0.35}

\newcommand{\vASMladSone}{0.66}

\newcommand{\vASMladStwothree}{0.58}
\newcommand{\vBLEHdyes}{0.17}
\newcommand{\vBLEHdyesRtwo}{0.79}

\newcommand{\vBLEHlad}{0.32}

\newcommand{\vBSMdyesRtwo}{0.82}

\newcommand{\vBSMlad}{0.49}

\newcommand{\vLEHLexc}{0.30}

\newcommand{\vLEHLlarge}{0.15}

\newcommand{\vLEHdyes}{0.29}

\newcommand{\vLEHlad}{0.36}
\newcommand{\vLEHlarge}{0.17}

\newcommand{\vLEHrand}{0.18}

\newcommand{\vMSANIrand}{0.41}

\newcommand{\vOMNIFTdyes}{0.88}

\newcommand{\vOMNIFTlad}{0.77}
\newcommand{\vOMNIFTlarge}{0.62}

\newcommand{\vOMNIFTrand}{0.60}

\newcommand{\vOMNIrand}{0.45}

\newcommand{\vSMdyes}{0.35}

\newcommand{\vSMlad}{0.61}
\newcommand{\vSMlarge}{0.19}

\newcommand{\vSMrand}{0.17}

\newcommand{\vXMdyes}{0.96}

\newcommand{\vXMlad}{0.77}
\newcommand{\vXMlarge}{0.21}

\newcommand{\vXMrand}{0.17}

\title{Toward universal excited-state machine-learned potentials: a local excitation Hamiltonian for size consistency and transferability}
\author{Pascal Friederich$^{1,2}$\\[1ex]
  {\small $^1$Institute of Nanotechnology, Karlsruhe Institute of Technology,}\\
  {\small Kaiserstr.~12, 76131 Karlsruhe, Germany}\\[0.5ex]
  {\small $^2$Institute for Anthropomatics and Robotics, Karlsruhe Institute of Technology,}\\
  {\small Kaiserstr.~12, 76131 Karlsruhe, Germany}\\[0.5ex]
  {\small \href{mailto:pascal.friederich@kit.edu}{pascal.friederich@kit.edu}, ORCID \href{https://orcid.org/0000-0003-4465-1465}{0000-0003-4465-1465}}}
\date{}

\begin{document}
\makeatletter\@namedef{b@achemso-control}{}\makeatother\nocite{achemso-control}  %
\maketitle

\begin{abstract}
Universal machine-learned interatomic potentials could make excited-state simulations across chemical space routine, but excitation energies, unlike ground-state energies, are not sums of atomic contributions. The excited states of noninteracting molecules are the union of their states, which pooled multistate readouts reproduce at best for the lowest state. We introduce LEH-MACE, whose excitation energies are the eigenvalues of a local excitation Hamiltonian assembled from atom-centred blocks, so this union holds exactly. We generated TD-DFT energies and forces for about 300,000 equilibrium, thermal, and excited-state geometries of over 100,000 molecules. LEH-MACE's advantage over direct readouts grew with the distance from the training data and was largest above $S_1$, with about half their $S_2$ and $S_3$ errors for long conjugated oligomers. For molecule pairs, only LEH-MACE kept its single-molecule accuracy, while a softmin readout remained superior for $S_1$ dynamics of larger chromophores. Exact size consistency is thus one step toward such potentials.
\end{abstract}

\section*{Introduction}
\label{sec:intro}
Machine-learned interatomic potentials (MLIPs) reach the accuracy of their quantum-chemical reference methods at a small fraction of the cost and have become standard tools for simulations in the electronic ground state~\cite{friederich2021natmat}. For electronically excited states, MLIPs can make photodynamics simulations of molecules orders of magnitude cheaper than on-the-fly electronic-structure calculations~\cite{westermayr2021review,dral2021nrc,li2022acr}. Since the first nonadiabatic dynamics simulations with MLIPs~\cite{hu2018jpcl,dral2018jpcl,chen2018jpcl}, excited-state MLIPs have enabled photodynamics on nanosecond time scales~\cite{westermayr2019chemsci,li2021znsh}, mechanistic studies of photochemical reactions~\cite{li2021cubanes}, the virtual screening of photoswitches~\cite{axelrod2022,axelrod2023halflives}, and studies of the photodegradation of materials for organic light-emitting diodes~\cite{kumar2026oled}, with architectures that also predict the couplings between states~\cite{westermayr2020schnarc,mausenberger2024spainn}. Most of these models are trained for one molecule or a small family of related molecules. A diabatic neural network trained on azobenzene derivatives transferred to azobenzenes with substitution patterns outside its training set and predicted their photoisomerization quantum yields~\cite{axelrod2022}, and models that transfer across chemical space more broadly have been proposed only recently~\cite{westermayr2020uv,martyka2025msani,barrett2025xmace,martyka2026omni}. Such a model has to solve two problems that ground-state MLIPs do not face.

The first problem is the structure of the target. Ground-state energies are, to a good approximation, sums of local contributions, which makes atom-centred MLIPs transferable to larger systems. Excitation energies are not additive. For two noninteracting fragments A and B, the locally excited states of A+B are the union of the excited states of A and of B, so that $\Delta E_1(\text{A}+\text{B})=\min[\Delta E_1(\text{A}),\Delta E_1(\text{B})]$ and two identical fragments give exactly degenerate pairs of states. A multistate readout that sums atomic contributions, as used in many excited-state MLIPs~\cite{westermayr2020schnarc,martyka2025msani}, produces excitation energies that grow with system size. Mean pooling is intensive but averages the fragments. Extremal pooling recovers the lowest excitation~\cite{malosso2026extremal}, and set-based readouts~\cite{barrett2025xmace} and fragment corrections~\cite{martyka2026omni} have been proposed, but none of these readouts reproduces the union of the spectra for the whole low-lying spectrum.

Hamiltonian readouts have this property by construction, because the spectrum of a block-diagonal matrix is the union of the spectra of its blocks. Learned Hamiltonians have been used to predict orbital energies from a fixed-size pseudo-Hamiltonian~\cite{westermayr2021schnetH}, excitation energies of hydrocarbons from an atom-centred minimal-basis one-electron Hamiltonian followed by a simplified Tamm--Dancoff calculation~\cite{cignoni2024halex}, diabatic two-state Hamiltonians for photoswitches~\cite{axelrod2022}, exciton Hamiltonians of molecular aggregates built from physically defined local and charge-transfer states~\cite{ren2025exciton}, and latent Hamiltonians for the excited states of individual molecules~\cite{juergens2026lush}. Whether a latent excitation Hamiltonian, assembled from local atom-centred blocks and trained only on adiabatic excitation energies and forces, transfers across chemical space and to larger molecules has not been tested.

The second problem is data. Large excited-state databases such as QM8~\cite{ramakrishnan2015qm8}, QCDGE~\cite{zhu2024qcdge}, and the data used for OMNI-P2x~\cite{martyka2026omni} cover hundreds of thousands of molecules, but mostly at equilibrium geometries and without excited-state forces. Datasets for excited-state dynamics, such as SHNITSEL~\cite{curth2025shnitsel} and WS22~\cite{pinheiro2023ws22}, contain many geometries of a few molecules. Neither kind of dataset is suited to train and test a model that transfers across chemical space and remains accurate away from equilibrium, where excited-state dynamics takes place.

Here we address both problems. We introduce LEH-MACE, a local excitation Hamiltonian (LEH) predicted by a MACE~\cite{batatia2022mace} encoder: onsite blocks for every atom and coupling blocks for atom pairs within a cutoff form a symmetric matrix whose sorted eigenvalues are the singlet excitation energies (Fig.~\ref{fig:schematic}). The predicted spectrum of separated molecules is therefore exactly the union of their spectra. We also present a TD-$\omega$B97X-D/6-31G* dataset of \nGeomAll{} geometries of \nMolAll{} molecules with ground- and excited-state energies and forces at equilibrium, thermal, and excited-state-sampled geometries, together with test sets of larger molecules, conjugated dyes, and ladder molecules with up to 30 heavy atoms. Of the readouts we compared (direct multistate readouts on the same encoder, X-MACE~\cite{barrett2025xmace}, and OMNI-P2x~\cite{martyka2026omni}), only LEH is exactly size consistent. Its advantage over the other readouts grew with the distance of the test molecules from the training data, from larger molecules to conjugated dyes and ladder molecules, and it was largest for pairs of molecules, where only LEH kept its single-molecule accuracy. For molecules similar to the training molecules, the direct readouts were slightly more accurate, and for the lowest excited state, the direct softmin readout remained competitive also beyond them. An active-learning loop along model-driven $S_1$ dynamics added \nALGeom{} geometries in twisted and low-gap regions, on which we trained production models for excited-state dynamics.

\section*{Results}
\label{sec:results}

\subsection*{A local excitation Hamiltonian}
\begin{figure}[tbp]
  \centering
  \includegraphics[width=\textwidth]{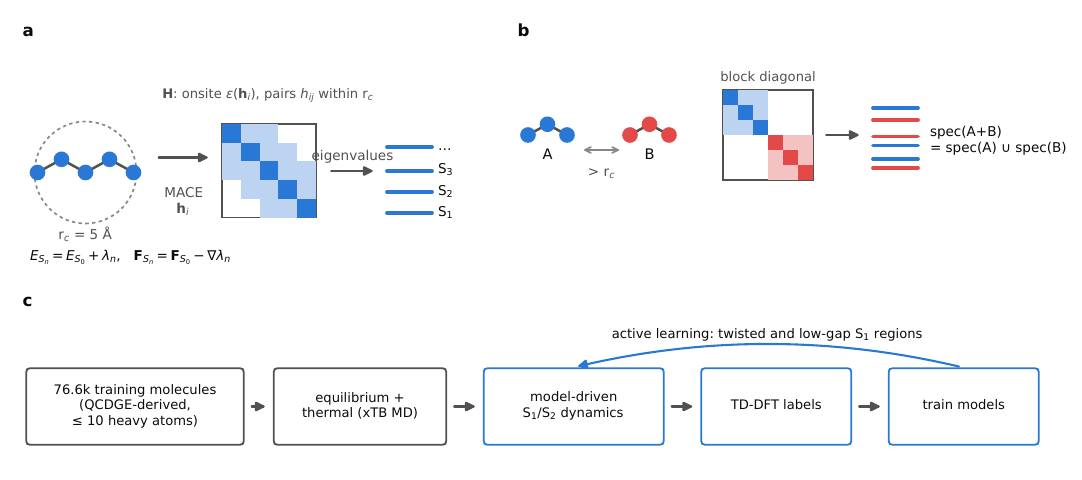}
  \caption{\textbf{Local excitation Hamiltonian and data.} \textbf{a}, A MACE encoder computes atom features $\mathbf{h}_i$ within the cutoff $r_c=5$\,\AA. Onsite blocks and pair blocks form a symmetric matrix $\mathbf{H}$, whose sorted eigenvalues are the excitation energies $S_1, S_2, \dots$ \textbf{b}, Two molecules A and B further apart than $r_c$. $\mathbf{H}$ is block diagonal, and its spectrum is the union of the spectra of A and B. \textbf{c}, Generation of the training data. QCDGE-derived molecules at equilibrium and thermal geometries and geometries from model-driven $S_1$ and $S_2$ dynamics are labelled with TD-$\omega$B97X-D/6-31G* energies and forces ($S_0$--$S_3$, ten singlet excitation energies). An active-learning loop repeats the excited-state sampling for the production models.}
  \label{fig:schematic}
\end{figure}

LEH replaces the multistate readout of an MLIP by a learned symmetric matrix (Fig.~\ref{fig:schematic}a). A MACE encoder computes features $\mathbf{h}_i$ for every atom. An onsite network maps them to one diagonal element per atom, and a pair network gives a coupling for every pair of atoms closer than $r_c=5$\,\AA. The sorted eigenvalues of this matrix are the singlet excitation energies, the ground-state energy is a sum of atomic energies, and the excited-state forces follow from the derivatives of the eigenvalues (Methods). The matrix grows with the molecule, so a larger system has more low-lying states. For two molecules further apart than $r_c$, the matrix is block diagonal, and its spectrum is exactly the union of the two spectra (Fig.~\ref{fig:schematic}b). Readouts that pool over atoms, by summing, averaging, or taking a soft minimum, and models built on per-state atomic sums do not have this property.

\subsection*{An off-equilibrium excited-state dataset}
\begin{figure}[tbp]
  \centering
  \includegraphics[width=\textwidth]{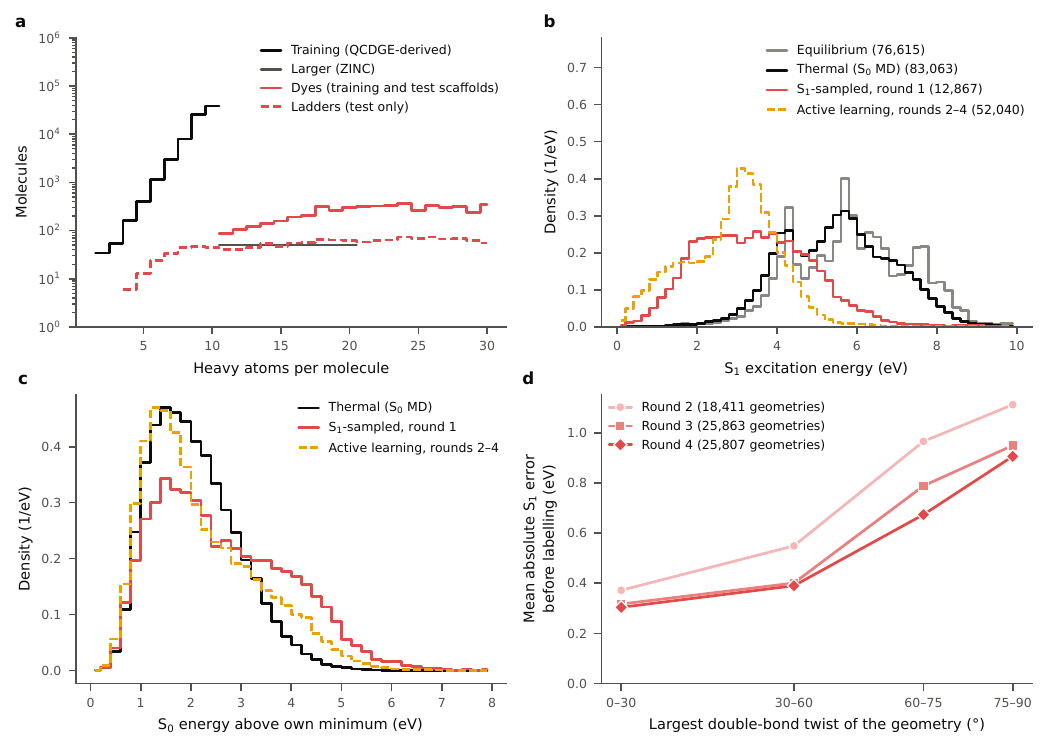}
  \caption{\textbf{Dataset.} \textbf{a}, Number of molecules versus the number of heavy atoms for the QCDGE-derived training molecules, the larger test molecules, the dyes (training and test scaffolds, the training scaffolds are part of the production training set only), and the ladder molecules (test only). \textbf{b}, Distribution of the $S_1$ excitation energy of the production training geometries by geometry type (number of geometries in brackets). \textbf{c}, Distribution of the $S_0$ energy above the minimum of the same molecule. \textbf{d}, Mean absolute $S_1$ error of the sampling models of active-learning rounds 2--4 on the geometries they selected, before labelling, versus the largest double-bond twist of the geometry.}
  \label{fig:dataset}
\end{figure}

Excited-state dynamics leaves the equilibrium geometry within femtoseconds, so a transferable model needs excited-state forces at distorted geometries of many molecules. We computed TD-$\omega$B97X-D/6-31G* ground- and excited-state energies and forces ($S_0$--$S_3$, ten singlets) for \nGeomAll{} geometries of \nMolAll{} molecules, \nGeomFex{} of them with excited-state forces (Figs.~\ref{fig:schematic}c and~\ref{fig:dataset}, Supplementary Note~1, Supplementary Fig.~1). Most molecules are derived from QCDGE~\cite{zhu2024qcdge}. Each of them contributes its QCDGE minimum and thermal geometries from ground-state xTB dynamics at 300--1000\,K. Another \nGeomExcAll{} geometries come from model-driven $S_1$ and $S_2$ dynamics, from one round of sampling for the benchmark models and three active-learning rounds for the production models (Methods). The thermal geometries reach several eV above the minimum in $S_0$ but rarely small $S_1$ excitation energies, which the excited-state-sampled geometries cover (Fig.~\ref{fig:dataset}b,c). We trained the benchmark models of the following sections on the first version of the dataset, \nTrainGeom{} geometries of \nTrainMol{} molecules with up to ten heavy atoms. The test sets contain random and unseen-scaffold molecules of the same size, \nLargeMol{} larger molecules with 11--20 heavy atoms, \nDyes{} conjugated dyes, and \nLadders{} ladder molecules with up to 30 heavy atoms (Fig.~\ref{fig:dataset}a, Supplementary Fig.~2), none of which the benchmark models saw in training.

\subsection*{Accuracy on held-out molecules}
\begin{figure}[tbp]
  \centering
  \includegraphics[width=\textwidth]{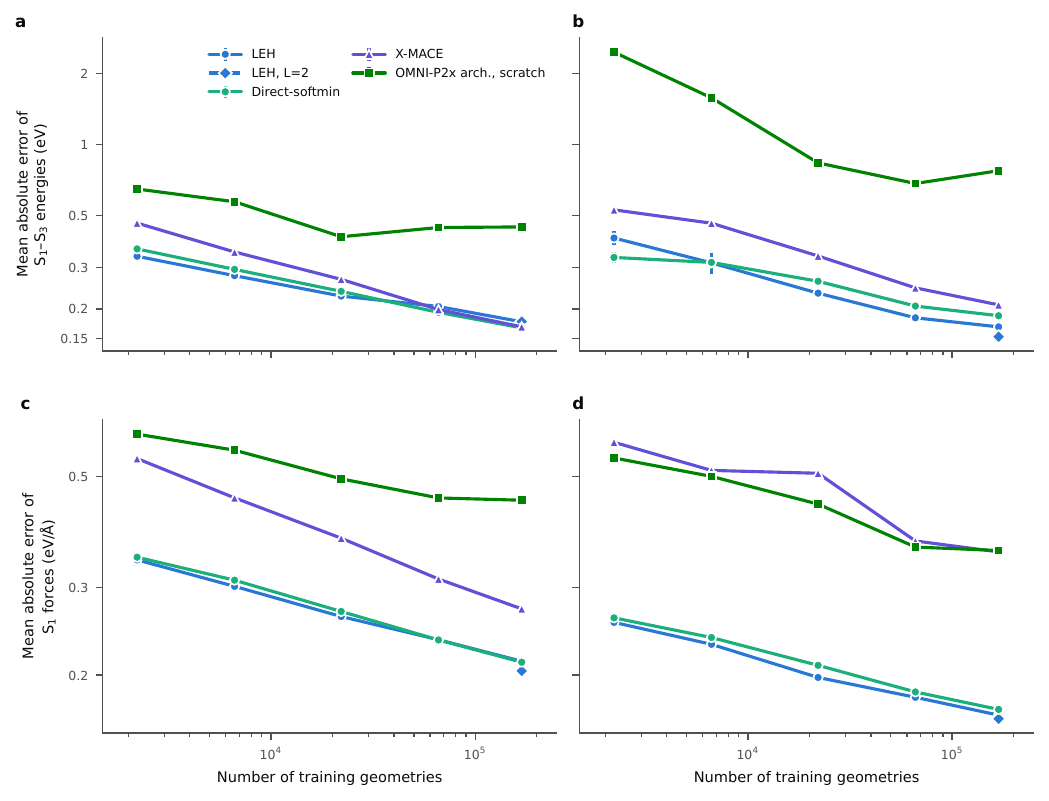}
  \caption{\textbf{Learning curves.} Mean absolute error of the $S_1$--$S_3$ excitation energies (\textbf{a}, \textbf{b}) and of the $S_1$ forces (\textbf{c}, \textbf{d}) versus the number of training geometries, for random test molecules with up to 10 heavy atoms at thermal geometries (\textbf{a}, \textbf{c}) and larger test molecules with 11--20 heavy atoms (\textbf{b}, \textbf{d}). The training sets contain 1k--76.6k molecules with about 2.2 geometries per molecule. Markers show seed means and bars seed standard deviations. LEH, $L=2$: encoder with higher angular resolution, full training set only. OMNI-P2x arch.: the OMNI-P2x architecture trained from scratch on our data. MS-ANI and the pretrained and fine-tuned OMNI-P2x are shown in Supplementary Fig.~3. All models use the same splits and test sets.}
  \label{fig:lc}
\end{figure}

For test molecules similar to the training molecules, LEH, the direct softmin readout on the same encoder, and X-MACE~\cite{barrett2025xmace} reached similar accuracy with the full training set, with $S_1$--$S_3$ errors of \vLEHrand{}\,eV, \vSMrand{}\,eV, and \vXMrand{}\,eV for random molecules at thermal geometries (Fig.~\ref{fig:lc}a, Table~\ref{tab:benchmark}). With fewer training data, LEH was the most accurate model, and for larger molecules it had the lowest error from 10k training molecules on (Fig.~\ref{fig:lc}b). With the full training set, the direct readouts had slightly lower errors for unseen scaffolds and $S_1$-sampled geometries. An encoder with higher angular resolution ($L=2$, one seed) reduced the error of LEH for larger molecules to \vLEHLlarge{}\,eV, the lowest of all models, and for $S_1$-sampled geometries to \vLEHLexc{}\,eV (Table~\ref{tab:benchmark}). The OMNI-P2x architecture trained from scratch on the same data reached \vOMNIrand{}\,eV for random molecules, and MS-ANI, the same architecture without the level-of-theory input, \vMSANIrand{}\,eV. The pretrained OMNI-P2x model fine-tuned with the protocol of its authors was less accurate for these molecules (\vOMNIFTrand{}\,eV) but more accurate than training from scratch for larger and conjugated molecules (Supplementary Note~3). The $S_1$ forces differed more between the models than the energies. LEH and the direct softmin readout predicted them with errors of \fLEHrand{}\,eV/\AA{} and \fSMrand{}\,eV/\AA, and X-MACE with \fXMrand{}\,eV/\AA{} (Fig.~\ref{fig:lc}c,d). The ground-state energies of LEH and Direct-softmin had errors of \eLEHrand{}\,meV per atom and \eSMrand{}\,meV per atom for random molecules and \eLEHlarge{}\,meV per atom and \eSMlarge{}\,meV per atom for larger molecules. X-MACE had the lowest ground-state error for random molecules (\eXMrand{}\,meV per atom) but three times the error of LEH for larger molecules (\eXMlarge{}\,meV per atom, Table~\ref{tab:benchmark}). At the equilibrium geometries, where the training data contain no excited-state forces, LEH predicted the $S_1$ forces, which set the initial direction of an excited-state relaxation, with errors of \fcLEHrand{}\,eV/\AA{} and \fcLEHlarge{}\,eV/\AA{} for random and larger molecules ($R^2=\fcLEHrtwo$, X-MACE: \fcXMrand{}\,eV/\AA{} and \fcXMlarge{}\,eV/\AA, Supplementary Note~10). For LEH, Direct-softmin, and X-MACE, all errors decreased with the number of training geometries and did not reach a plateau with the full training set, whereas the errors of the OMNI-P2x architecture changed little beyond 10k training molecules. At a fixed number of training geometries, the accuracy did not depend on whether they came from more molecules or from more geometries per molecule (Supplementary Note~8).

\begin{table}[tbp]
  \centering
  \caption{\textbf{Errors of the models trained on the full training set.} Mean absolute errors of the $S_1$--$S_3$ excitation energies for random, unseen-scaffold, $S_1$-sampled, and larger test molecules, of the $S_1$ forces, and of the $S_0$ energy. Mean $\pm$ standard deviation over seeds where available. Bold: lowest error per column. All targets with $R^2$ and the decomposition of the $S_1$ energy error into $S_0$ and excitation-energy errors are listed in Supplementary Table~2.}
  \label{tab:benchmark}
  \resizebox{\textwidth}{!}{\begin{tabular}{lcccccccc}
\toprule
 & \multicolumn{4}{c}{$S_1$--$S_3$ MAE (meV)} & \multicolumn{2}{c}{$S_1$ forces (meV/\AA)} & \multicolumn{2}{c}{$S_0$ (meV/atom)} \\
\cmidrule(lr){2-5}\cmidrule(lr){6-7}\cmidrule(lr){8-9}
Model & random & scaffold & $S_1$-sampl. & larger & random & larger & random & larger \\
\midrule
LEH & 177$\pm$2 & 207$\pm$5 & 326$\pm$9 & 168$\pm$2 & 213$\pm$1 & 166$\pm$1 & 5.1$\pm{<}0.1$ & 3.9$\pm$0.2 \\
LEH, $L=2$ & 177 & 207 & 300 & \textbf{153} & \textbf{204} & \textbf{164} & 4.6 & 3.9 \\
Direct-softmin & \textbf{167$\pm{<}1$} & 189$\pm{<}1$ & 279$\pm$1 & 187$\pm$2 & 212$\pm$1 & 171$\pm$2 & 4.9$\pm{<}0.1$ & \textbf{3.7$\pm$0.1} \\
X-MACE & 168 & \textbf{186} & \textbf{268} & 208 & 272 & 353 & \textbf{4.0} & 12.1 \\
OMNI-P2x arch. & 447 & 462 & 421 & 774 & 448 & 356 & 10.7 & 8.2 \\
\bottomrule
\end{tabular}
}
\end{table}

\subsection*{Transfer beyond the training distribution}
\begin{figure}[tbp]
  \centering
  \includegraphics[width=\textwidth]{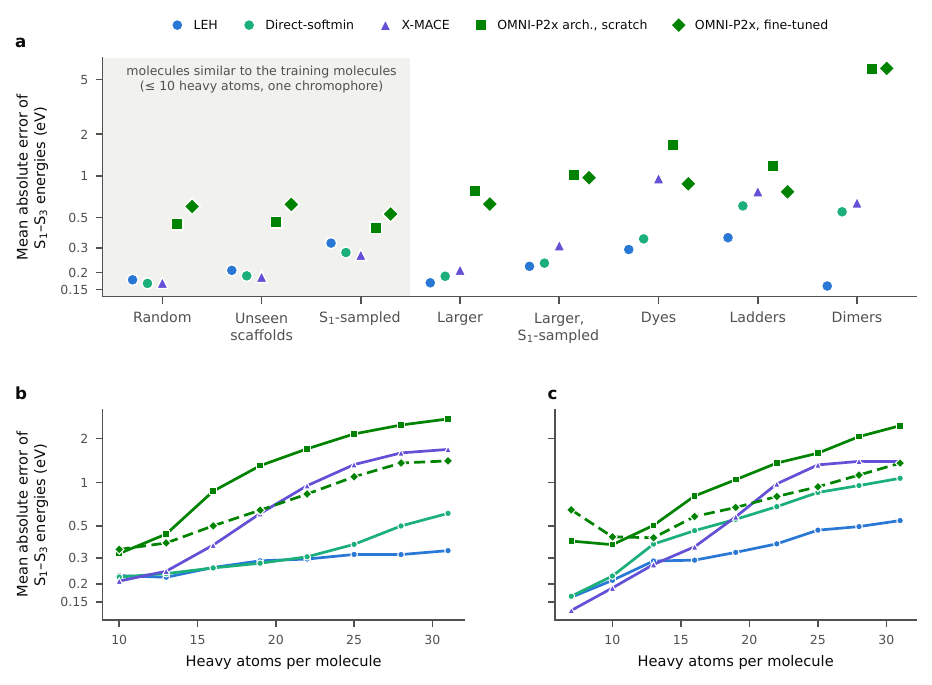}
  \caption{\textbf{Transferability.} \textbf{a}, Mean absolute error of the $S_1$--$S_3$ excitation energies of the models trained on the full training set, for test sets ordered by their distance from the training distribution. Shaded: test molecules similar to the training molecules. Dyes: all \nDyesNoS{} sulfur-free dyes at their equilibrium geometries, none of which these models saw in training. Ladders: all sulfur-free ladder molecules at their equilibrium geometries. Dimers: two non-bonded test molecules at minimum distances of 2.5--15\,\AA{}, compared with TD-$\omega$B97X-D dimer spectra (Fig.~\ref{fig:separability}). \textbf{b}, \textbf{c}, Mean absolute error of the $S_1$--$S_3$ excitation energies versus the number of heavy atoms for dyes (\textbf{b}) and ladder molecules (\textbf{c}) at their equilibrium geometries, in bins of three heavy atoms. Errors relative to LEH for $S_1$--$S_3$ and for $S_1$ alone are shown in Supplementary Fig.~6, and MS-ANI, the pretrained OMNI-P2x, and the adapted fine-tuning protocol in Supplementary Fig.~4. OMNI-P2x arch.: OMNI-P2x architecture trained from scratch on our data. OMNI-P2x, fine-tuned: pretrained OMNI-P2x fine-tuned on our data with the protocol of its authors.}
  \label{fig:transfer}
\end{figure}

LEH and the direct readouts differed most when the test molecules differed from the training molecules (Fig.~\ref{fig:transfer}a). The direct softmin readout and X-MACE were slightly more accurate than LEH for molecules similar to the training molecules, but their errors grew faster with the distance from the training distribution. For larger molecules, LEH reached \vLEHlarge{}\,eV (Direct-softmin \vSMlarge{}\,eV, X-MACE \vXMlarge{}\,eV), for the \nDyesNoS{} sulfur-free dyes, none of which these models saw in training, \vLEHdyes{}\,eV (\vSMdyes{}\,eV and \vXMdyes{}\,eV), and for ladder molecules \vLEHlad{}\,eV (\vSMlad{}\,eV and \vXMlad{}\,eV). The difference increased with molecule size (Fig.~\ref{fig:transfer}b,c). The error of X-MACE increased most among these three models, because X-MACE decodes all excited-state energies from one sum of atomic features over the whole molecule, and this sum leaves the range seen in training for larger molecules (Methods, Supplementary Note~9). The OMNI-P2x architecture, which predicts each state as a sum of atomic energies, had the strongest size dependence. Fine-tuning the pretrained OMNI-P2x model reduced its errors for larger molecules, dyes, and ladder molecules to \vOMNIFTlarge{}\,eV, \vOMNIFTdyes{}\,eV, and \vOMNIFTlad{}\,eV, which are still two to four times those of LEH.

Most of the advantage of LEH lies in the states above $S_1$. For $S_2$ and $S_3$, LEH had errors of \vALEHladStwothree{}\,eV for the ladder molecules and \vALEHdyesStwothree{}\,eV for the held-out dyes, compared with \vASMladStwothree{}\,eV and \vASMdyesStwothree{}\,eV for the softmin readout. For $S_1$ alone, the two readouts had similar errors (ladder molecules: \vALEHladSone{}\,eV and \vASMladSone{}\,eV, held-out dyes: \vALEHdyesSone{}\,eV and \vASMdyesSone{}\,eV, Table~\ref{tab:conj}, Supplementary Fig.~6). This is consistent with a soft minimum over atomic contributions, which targets the lowest excitation but not the states above it. The production models, which we trained additionally on the active-learning data (Methods), showed the same pattern for the ladder molecules, with $S_2$--$S_3$ errors of \pLEHladStwothree{}\,eV versus \pSMladStwothree{}\,eV and $S_1$ errors of \pLEHladSone{}\,eV versus \pSMladSone{}\,eV (LEH versus softmin readout, committee predictions, Table~\ref{tab:conj}).

The benchmark models captured little of the variation of $S_1$ between the held-out dye scaffolds ($R^2$ of \vALEHdyesRtwo{} for LEH and \vASMdyesRtwo{} for the softmin readout, negative for X-MACE and both OMNI-P2x models, Table~\ref{tab:conj}). Adding the training dye scaffolds to the training data (dye-augmented models) reduced the errors for held-out dye scaffolds to \vBLEHdyes{}\,eV for both LEH and the softmin readout and raised the $R^2$ of $S_1$ to \vBLEHdyesRtwo{} and \vBSMdyesRtwo{}, without changing the accuracy for small molecules. For the ladder molecules, which remained unseen, LEH was more accurate (\vBLEHlad{}\,eV versus \vBSMlad{}\,eV, Table~\ref{tab:conj}).

\begin{table}[tbp]
  \centering
  \caption{\textbf{Conjugated molecules.} Mean absolute errors of the $S_1$--$S_3$ and $S_1$ excitation energies and $R^2$ of $S_1$ for held-out dye scaffolds and ladder molecules (sulfur-free) at their equilibrium geometries. The benchmark models were trained on molecules with up to ten heavy atoms, the dye-augmented models additionally on the training dye scaffolds, and the production models on the data of the dye-augmented models and all active-learning geometries with the weighted loss (Methods). No model was trained on ladder molecules. Mean $\pm$ standard deviation over seeds (two for the benchmark and dye-augmented models where available, four for the production models). Committee: seed-averaged prediction. OMNI-P2x, fine-tuned: average of the three fine-tuned ensemble members. Bold: lowest error or highest $R^2$ per column within each training-data group (for the production models, seed means and committees separately). Thermal geometries are listed in Supplementary Tables~7 and~11.}
  \label{tab:conj}
  \resizebox{\textwidth}{!}{\begin{tabular}{llcccccc}
\toprule
 & & \multicolumn{3}{c}{held-out dye scaffolds} & \multicolumn{3}{c}{ladder molecules} \\
\cmidrule(lr){3-5}\cmidrule(lr){6-8}
Training data & Model & $S_1$--$S_3$ MAE & $S_1$ MAE & $S_1$ $R^2$ & $S_1$--$S_3$ MAE & $S_1$ MAE & $S_1$ $R^2$ \\
 & & (meV) & (meV) & & (meV) & (meV) & \\
\midrule
\multirow{5}{*}{benchmark} & LEH & \textbf{292$\pm$17} & 406$\pm$35 & \textbf{0.17$\pm$0.02} & \textbf{357$\pm$12} & \textbf{596$\pm$10} & \textbf{0.48$\pm$0.01} \\
 & Direct-softmin & 365$\pm$38 & \textbf{392$\pm$42} & 0.04$\pm$0.33 & 606$\pm$11 & 660$\pm$6 & 0.31$\pm$0.01 \\
 & X-MACE & 1125 & 1543 & -8.71 & 772 & 891 & -0.54 \\
 & OMNI-P2x arch. & 1821 & 1336 & -6.51 & 1182 & 1199 & -1.62 \\
 & OMNI-P2x, fine-tuned & 976 & 657 & -1.16 & 767 & 679 & 0.21 \\
\midrule
\multirow{2}{*}{+ training dyes} & LEH & \textbf{171$\pm$15} & 207$\pm$36 & 0.79$\pm$0.07 & \textbf{324$\pm$76} & 532$\pm$163 & 0.51$\pm$0.25 \\
 & Direct-softmin & \textbf{171$\pm$7} & \textbf{188$\pm$5} & \textbf{0.82$\pm{<}0.01$} & 492$\pm$14 & \textbf{492$\pm{<}1$} & \textbf{0.57$\pm$0.01} \\
\midrule
\multirow{4}{*}{\shortstack[l]{production: + training dyes\\+ active learning}} & LEH & 171$\pm$7 & 192$\pm$6 & 0.81$\pm$0.02 & \textbf{283$\pm$58} & 442$\pm$141 & 0.64$\pm$0.18 \\
 & Direct-softmin & \textbf{160$\pm$11} & \textbf{173$\pm$13} & \textbf{0.86$\pm$0.02} & 313$\pm$28 & \textbf{325$\pm$26} & \textbf{0.81$\pm$0.02} \\
 & LEH, committee & 152 & 172 & 0.85 & \textbf{266} & 434 & 0.67 \\
 & Direct-softmin, committee & \textbf{148} & \textbf{164} & \textbf{0.88} & 307 & \textbf{321} & \textbf{0.82} \\
\bottomrule
\end{tabular}
}
\end{table}

\subsection*{Size consistency and noninteracting molecules}
\begin{figure}[tbp]
  \centering
  \includegraphics[width=\textwidth]{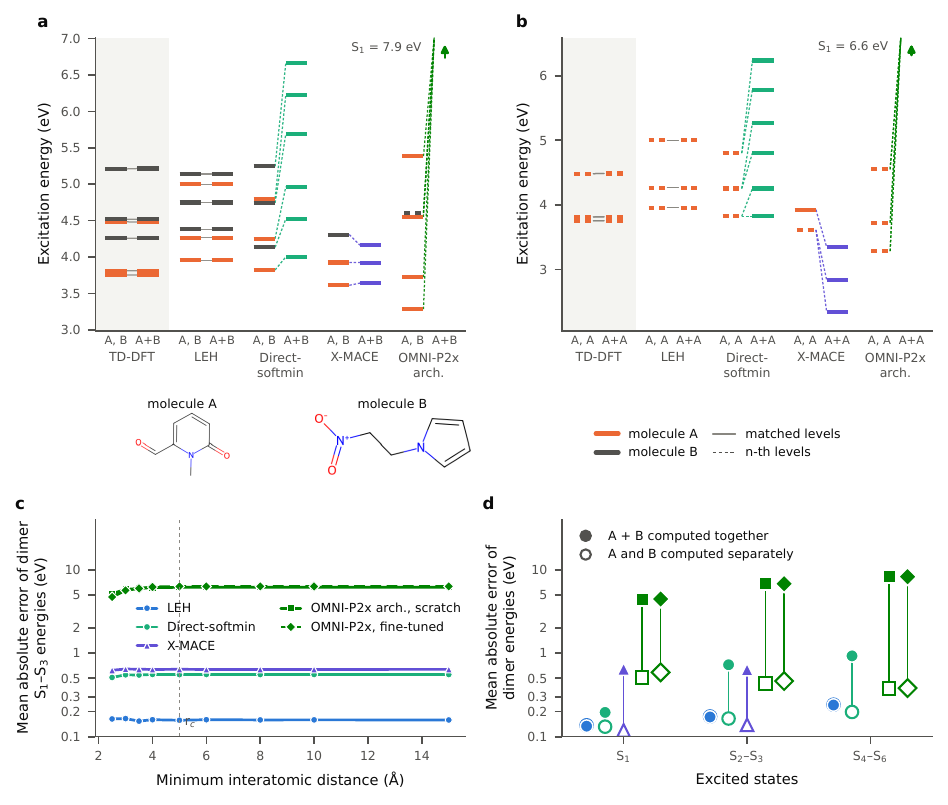}
  \caption{\textbf{Separability.} \textbf{a}, Excitation energies ($S_1$--$S_6$) of two different molecules A and B at a minimum interatomic distance of 10\,\AA, beyond the cutoff $r_c=5$\,\AA. For TD-$\omega$B97X-D and each model, the left column (A, B) shows the levels of A and B computed separately, coloured by molecule, and the right column (A+B) the levels of the pair computed together. Solid lines connect each level of the pair with the matching level of the separate molecules (TD-$\omega$B97X-D and LEH, dimer levels coloured by the molecule they belong to). Dashed lines connect the $n$-th levels of the two columns (pooled readouts, dimer levels in the model colour). X-MACE predicts $S_1$--$S_3$ only. Arrows mark levels above the plotted range. \textbf{b}, The same for two identical molecules (A + A, with A from panel \textbf{a}). Degenerate levels are drawn side by side. The pairs in \textbf{a} and \textbf{b} have the median and a near-median (fifth of ten) Direct-softmin error at 10\,\AA. LEH and Direct-softmin: seed 0. \textbf{c}, Mean absolute error of the dimer $S_1$--$S_3$ excitation energies with respect to TD-$\omega$B97X-D versus the minimum interatomic distance (\nDimerPairs{} dimers of ten different and ten identical pairs). The dashed line marks $r_c$. \textbf{d}, Mean absolute error of the dimer excitation energies for $S_1$, $S_2$--$S_3$, and $S_4$--$S_6$, for the pair computed together (filled) and for the union of the predictions for the two molecules computed separately (open). In \textbf{c} and \textbf{d}, models trained on the full training set, seed mean where two seeds exist. Readouts on the same encoder are compared in Supplementary Fig.~7.}
  \label{fig:separability}
\end{figure}

Only LEH kept its single-molecule accuracy when a second molecule was added (Fig.~\ref{fig:transfer}a). For two molecules 10\,\AA{} apart, the levels that LEH predicted for the pair agreed within $10^{-14}$\,eV with the levels it predicted for the two molecules separately, and for two identical molecules every level appeared twice, as in TD-$\omega$B97X-D (Fig.~\ref{fig:separability}a,b). The pooled readouts shifted and spread the levels of the pair, and the per-state atomic sums of the OMNI-P2x architecture moved them up by several eV.

At short distances, the two molecules interact. TD-$\omega$B97X-D shifts the dimer excitation energies relative to the union of the monomer spectra by \tdshiftABshort{}\,meV and \tdshiftAAshort{}\,meV at 2.5\,\AA{} (different and identical molecules, the latter by excitonic splitting of degenerate states) and by only \tdshiftABfive{}--\tdshiftAAfive{}\,meV at 5\,\AA. LEH reproduced the order of magnitude of these short-range shifts and is exactly separable beyond its cutoff. Its error for the \nDimerPairs{} dimers (\dimLEHAB{}\,eV and \dimLEHAA{}\,eV for different and identical pairs) was therefore no larger than for the same molecules alone (\monoLEH{}\,eV, Fig.~\ref{fig:separability}c). The other models were about as accurate as LEH for the single molecules (Direct-softmin \monoSM{}\,eV, X-MACE \monoXM{}\,eV) but not for the dimers (\dimSM{}\,eV and \dimXM{}\,eV), at every distance up to 15\,\AA. Per-state atomic sums predicted the excitation energy of a dimer approximately as the sum of those of its monomers (OMNI-P2x architecture: \dimOMNI{}\,eV, fine-tuned OMNI-P2x: \dimOMNIFT{}\,eV).

For Direct-softmin, the dimer error came mainly from the states above $S_1$. Its dimer $S_1$ error was moderately larger than that of LEH (\sepSMone{}\,eV versus \sepLEHone{}\,eV), consistent with a soft minimum, which selects the lower of the two monomer excitations. Its $S_2$--$S_3$ and $S_4$--$S_6$ errors were several times larger than those of LEH (\sepSMtwothree{}\,eV versus \sepLEHtwothree{}\,eV and \sepSMfoursix{}\,eV versus \sepLEHfoursix{}\,eV, Fig.~\ref{fig:separability}d). The union of its own predictions for the two molecules computed separately would have been as accurate as LEH (\sepSMunion{}\,eV), so the error comes from the readout and not from the encoder. Readouts that pool by summing, averaging, or a soft minimum on the same encoder deviated from the union of their monomer spectra by 0.4--1.3\,eV at all distances (Supplementary Fig.~7).

\subsection*{Public benchmark}
On the public QCDGE dataset, we trained on equilibrium geometries of molecules with up to eight heavy atoms and tested on molecules with nine and ten heavy atoms (Table~\ref{tab:qcdge}). LEH and the softmin readout were equally accurate within the training size range (\qLEHval{}\,eV and \qSMval{}\,eV) and for nine heavy atoms. For ten heavy atoms, LEH was slightly more accurate for the three lowest states (\qLEHten{}\,eV versus \qSMten{}\,eV) and by a larger margin for all ten states (\qLEHtenall{}\,eV versus \qSMtenall{}\,eV). For ten-heavy-atom molecules with scaffolds absent from training, the two readouts were equally accurate (\qLEHtenscaf{}\,eV and \qSMtenscaf{}\,eV). Sum and mean pooling were less accurate. The error of the OMNI-P2x architecture increased from \qOMNIval{}\,eV to \qOMNIten{}\,eV, and that of X-MACE, which decodes all states from one sum over the atoms, from \qXMval{}\,eV to \qXMten{}\,eV ($R^2=\qXMtenrtwo$). The step from nine to ten heavy atoms is not only a step in size. In QCDGE, the molecules with up to nine heavy atoms are derived from QM9 and those with ten heavy atoms from GDB-11, so the ten-heavy-atom molecules also differ in their chemistry. LEH trained on a random split of the same size, in which \qRandNinePct{}\,\% of the training molecules have nine heavy atoms, reached \qLEHrndten{}\,eV on the ten-heavy-atom test molecules, close to the \qLEHten{}\,eV of the size split. Most of the error increase from nine to ten heavy atoms therefore comes from this change in chemistry and not from the larger size (Supplementary Note~13).

\begin{table}[tbp]
  \centering
  \caption{\textbf{QCDGE size extrapolation.} Mean absolute errors and $R^2$ of the excitation energies of models trained on QCDGE equilibrium geometries of molecules with up to eight heavy atoms (19,849 molecules, ten supervised excitation energies, no forces), for 5,000 test molecules each with nine and ten heavy atoms and for the ten-heavy-atom molecules with scaffolds absent from training. Mean $\pm$ standard deviation over two or three seeds (single runs without). $R^2$ per state, averaged over $S_1$--$S_3$. Bold: lowest error or highest $R^2$ per column. Random-split controls, $S_1$ errors, and the transition-dipole variant are listed in Supplementary Table~8.}
  \label{tab:qcdge}
  \IfFileExists{tables/table3_qcdge.tex}{\resizebox{\textwidth}{!}{\begin{tabular}{lcccccc}
\toprule
 & \multicolumn{4}{c}{$S_1$--$S_3$ MAE (meV)} & $S_1$--$S_{10}$ MAE (meV) & $S_1$--$S_3$ $R^2$ \\
\cmidrule(lr){2-5}
Heavy atoms & $\le 8$ (val.) & 9 & 10 & 10, new scaffolds & 10 & 10 \\
\midrule
LEH & \textbf{90$\pm$1} & \textbf{115$\pm$1} & \textbf{256$\pm$3} & 291$\pm$4 & \textbf{204$\pm$1} & \textbf{0.90$\pm{<}0.01$} \\
Direct-softmin & 92$\pm$1 & 117$\pm$3 & 265$\pm$5 & \textbf{285$\pm$7} & 229$\pm$6 & 0.89$\pm{<}0.01$ \\
Direct-sum & 103$\pm$1 & 139$\pm$1 & 293$\pm$3 & 297$\pm$17 & 268$\pm$10 & 0.88$\pm{<}0.01$ \\
Direct-mean & 102$\pm{<}1$ & 134$\pm$2 & 280$\pm$6 & 301$\pm$13 & 244$\pm$4 & 0.89$\pm{<}0.01$ \\
X-MACE & 182 & 260 & 849 & 682 & 903 & 0.23 \\
OMNI-P2x arch. & 178 & 240 & 511 & 494 & 517 & 0.64 \\
\bottomrule
\end{tabular}
}}{\todo{table pending}}
\end{table}

\subsection*{Excited-state dynamics}
\begin{figure}[tbp]
  \centering
  \includegraphics[width=\textwidth]{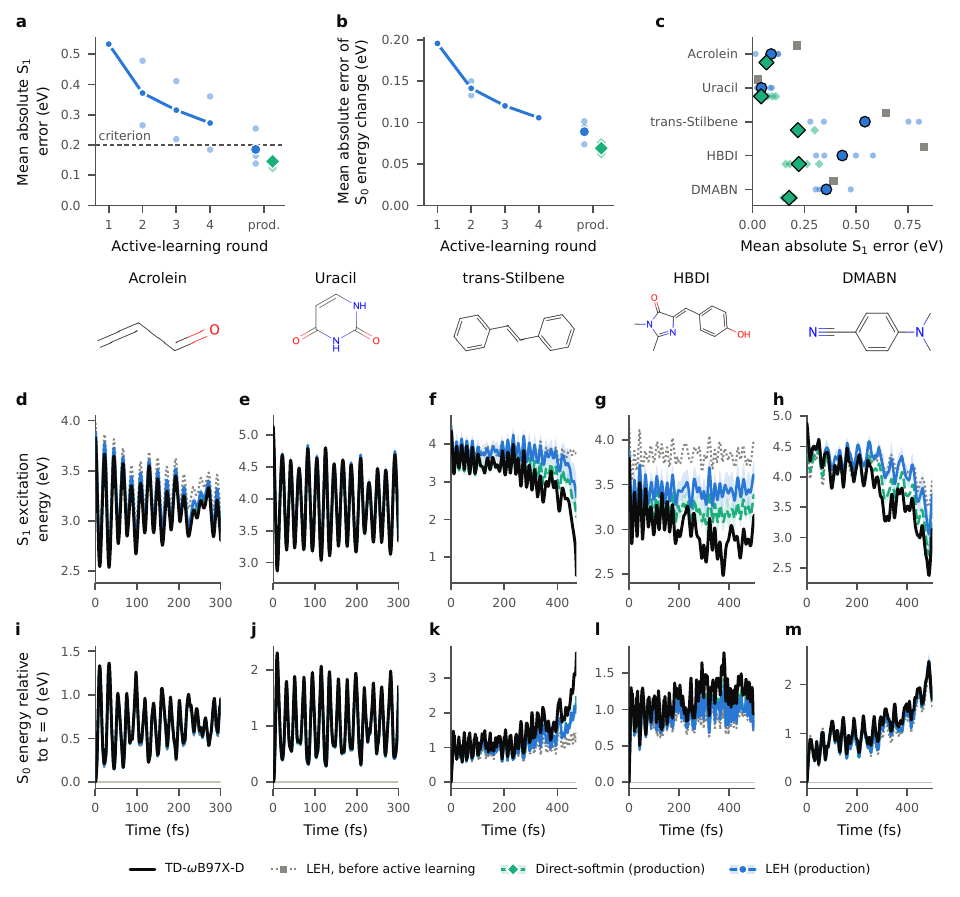}
  \caption{\textbf{Excited-state dynamics.} \textbf{a}, Mean absolute error of the $S_1$ excitation energy and \textbf{b}, of the change of the $S_0$ energy along the TD-$\omega$B97X-D $S_1$ trajectories of seven validation molecules, for the sampling models after each active-learning round (small markers: the two models, line: mean, round 1: benchmark LEH) and for the production models (small markers: four seeds, large markers: mean). The dashed line marks the 0.2\,eV threshold of the stopping criterion. \textbf{c}, Mean absolute $S_1$ error along the five test trajectories for the benchmark LEH before active learning and for the production LEH and Direct-softmin models (small markers: seeds, large outlined markers: committee). The structures of the test molecules are shown above \textbf{d}--\textbf{h}. \textbf{d}--\textbf{h}, $S_1$ excitation energy and \textbf{i}--\textbf{m}, $S_0$ energy relative to $t=0$ along the test trajectories, evaluated at the TD-$\omega$B97X-D geometries up to the frame where TD-DFT breaks down (acrolein and uracil: first 300\,fs). Production models: committee prediction (line) and range over the four seeds (band).}
  \label{fig:dynamics}
\end{figure}

Excited-state dynamics drives molecules into twisted geometries with small $S_1$ excitation energies, which thermal sampling and one round of excited-state sampling cover poorly. On the geometries selected in active-learning round 2, the $S_1$ error of the benchmark model increased from \alRtwoTwistLow{}\,eV below a double-bond twist of 30$^\circ$ to \alRtwoTwistHigh{}\,eV beyond 75$^\circ$ (Fig.~\ref{fig:dataset}d). Three active-learning rounds~\cite{zhou2025pal} along model-driven $S_1$ dynamics added \nALGeom{} geometries of \nALMol{} molecules (Methods). The error of the sampling models on the geometries they selected decreased from \alRtwoMAE{}\,eV to \alRfourMAE{}\,eV (Fig.~\ref{fig:dataset}d). Their mean $S_1$ error along the TD-$\omega$B97X-D $S_1$ trajectories of seven validation molecules, which never entered training, decreased from \alGapRone{}\,eV to \alGapRfour{}\,eV (Fig.~\ref{fig:dynamics}a, Supplementary Note~14), and the error of the change of their $S_0$ energy from \alRiseRone{}\,eV to \alRiseRfour{}\,eV (Fig.~\ref{fig:dynamics}b). Models trained from scratch on all data reached \alGapProdUniform{}\,eV on these trajectories, below the 0.2\,eV threshold at which the loop stops. For the production models, we used a loss that gives more weight to geometries with small $S_1$ excitation energies, which reduced the overestimation of $S_1$ near $S_1/S_0$ crossings (Methods). With four seeds each, the production LEH and Direct-softmin models reached \pLEHValTraj{}$\pm$\pLEHValTrajstd{}\,eV and \pSMValTraj{}$\pm$\pSMValTrajstd{}\,eV on the validation trajectories (Fig.~\ref{fig:dynamics}a) and errors of \pLEHValRise{}\,eV and \pSMValRise{}\,eV for the change of the $S_0$ energy (Fig.~\ref{fig:dynamics}b). Their own $S_1$ dynamics from the validation minima was stable, with a total-energy drift of at most \pStabDrift{}\,meV over 1\,ps and no bond changes.

We did not use the five test trajectories for any decision. Active learning reduced the mean $S_1$ error of LEH along these trajectories from \bLEHTestTraj{}\,eV for the benchmark model to \pLEHTestTrajcomm{}\,eV (Direct-softmin: \pSMTestTrajcomm{}\,eV, Fig.~\ref{fig:dynamics}c). Along an $S_1$ trajectory, the $S_1$ potential energy is the sum of the $S_0$ energy and the $S_1$ excitation energy. The excitation energy determines when a molecule reaches the crossing with $S_0$ (Fig.~\ref{fig:dynamics}d--h), and the $S_0$ energy rises as the molecule moves away from the ground-state minimum (Fig.~\ref{fig:dynamics}i--m). Both production models reproduced the $S_1$ excitation and $S_0$ energies of acrolein and uracil within \pSmallTrajMax{}\,eV. For the three larger chromophores, \textit{trans}-stilbene, HBDI, and DMABN, both models reproduced the increase of the $S_0$ energy by \pRiseRefMin{}--\pRiseRefMax{}\,eV within \pRiseErrMin{}--\pRiseErrMax{}\,eV. As the $S_1$ excitation energy decreased during relaxation and twisting, both models predicted it too high, on average by \pSMbiasMin{}--\pSMbiasMax{}\,eV (Direct-softmin) and \pLEHbiasMin{}--\pLEHbiasMax{}\,eV (LEH) for the committee predictions. All four LEH seeds overestimated it in the same way, so their disagreement does not indicate this error. At the same time, both models underestimated the rise of the $S_0$ energy, and the two errors largely cancelled (correlation coefficients between $-\pCorrMin{}$ and $-\pCorrMax{}$ for \textit{trans}-stilbene and HBDI). The $S_1$ potential energy, which determines the forces in $S_1$ dynamics, therefore had smaller errors than the excitation energy (Direct-softmin: \pSMpotMin{}--\pSMpotMax{}\,eV versus \pSMgapMin{}--\pSMgapMax{}\,eV, LEH: \pLEHpotMin{}--\pLEHpotMax{}\,eV versus \pLEHgapMin{}--\pLEHgapMax{}\,eV), but both models kept $S_1$ and $S_0$ too far apart near the twisted geometries. For $S_1$ dynamics of single molecules larger than the training molecules, Direct-softmin is therefore the more accurate of the two production models.

\section*{Discussion}
\label{sec:discussion}
A latent excitation Hamiltonian assembled from local blocks gives an excited-state MLIP a property that pooled multistate readouts lack: its spectrum for noninteracting molecules is exactly the union of the spectra of the molecules. For pairs of molecules, LEH was three to four times more accurate than every direct readout. For single molecules that are larger or more extended than the training molecules, LEH transferred better than the direct readouts, with lower errors mainly for the states above $S_1$. This advantage grew with the distance from the training distribution, where a model for photochemistry across chemical space is needed most. With up to 10k training molecules, LEH was the most accurate model, and its excited-state forces were as accurate as those of the best direct readout. With the full training set, the direct readouts had 0.01--0.06\,eV lower errors for molecules similar to the training molecules. The remaining errors of 0.15--0.2\,eV for the excitation energies of molecules similar to the training molecules are comparable to the error of TD-DFT with range-separated hybrids relative to high-level references, about 0.2--0.3\,eV~\cite{jacquemin2009,liang2022tddft,loos2020quest}.

For the lowest excited state of a single molecule, the direct softmin readout is the stronger model. A soft minimum over atomic contributions targets the lowest excitation~\cite{malosso2026extremal}, which is the state that governs $S_1$ dynamics. The production Direct-softmin model was more accurate than LEH for $S_1$ along the test trajectories of the larger chromophores and for $S_1$ of long ladder molecules. This difference does not come from the weighted loss, which made LEH less accurate along the trajectories, because the models trained with the uniform loss showed the same order (\uLEHTestTraj{}\,eV versus \uSMTestTraj{}\,eV along the test trajectories). In both cases, LEH predicted too high an $S_1$ excitation energy where extended conjugation or twisting lowers it. Our results presented here do not yet indicate why the eigenvalue readout learns this lowering less well. More latent channels per atom, longer-range couplings, and active learning on larger chromophores are possible approaches for improvement. The two readouts thus have different use-cases: LEH is suited on the path towards a universal MLIP, where the spectrum above $S_1$ or more than one molecule matters, for example for absorption spectra, photophysical processes that involve higher-lying excited states~\cite{he2026kinluv}, or chromophores in explicit solvent. Direct-softmin is suited for $S_1$ dynamics of single molecules.

\paragraph{Limitations and outlook.}
The latent Hamiltonian is size consistent only within the sector it represents, the union of local excitations of separated fragments. The short-range blocks do not represent charge-transfer states between fragments, double excitations, or long-range excitonic couplings. For the closed-shell organic molecules of this work, TD-$\omega$B97X-D shifts the dimer spectra by less than \tdshiftAAfive{}\,meV beyond 5\,\AA, well below the model error, but this will not hold for donor--acceptor complexes or molecular aggregates. A dipole--dipole coupling between distant channels would describe excitonic interactions in aggregates.
The eigenvalues determine the Hamiltonian only up to orthogonal transformations. The learned couplings and eigenvectors therefore have no diabatic meaning, and the model provides no nonadiabatic couplings. Unlike a pooled readout, LEH can nevertheless be trained with information on its eigenvectors. With a loss on the TD-DFT excitation populations of $S_1$--$S_3$ (10k training molecules), the weight shared by the $S_1$ eigenvector and the TD-DFT population on validation molecules increased from \popSvalBase{} to \popSvalPop{}, and the $S_1$--$S_3$ error of random test molecules decreased from \popRandBase{}\,eV to \popRandPop{}\,eV. For the ladder molecules, neither this agreement nor the $S_1$ error changed (\popLadBase{}\,eV and \popLadPop{}\,eV, Supplementary Note~7). Transition dipoles built from the eigenvectors give oscillator strengths, which were accurate within but not beyond the training size range in our tests (Supplementary Note~13).
The ground state is predicted by a separate extensive readout, to which the excitation energies are added. The topology of $S_1/S_0$ conical intersections is therefore not represented, whereas intersections between excited states arise from the eigenvalues of a smooth matrix. Linear-response TD-DFT itself describes $S_1/S_0$ intersections incorrectly~\cite{levine2006ci}, so an improvement would require other reference methods. A coupling of each eigenmode to the ground state, as in latent Hamiltonians that include the ground state~\cite{juergens2026lush}, would represent $S_1/S_0$ intersections without giving up separability, and spin-flip or multireference data would allow training it.
With one latent channel per atom, the model has as many states as atoms, which restricts diatomic and triatomic molecules to two or three states. More channels per atom remove this restriction.
LEH and the direct readouts predicted negative excitation energies for a few highly strained cage molecules of the scaffold test set. Soft bounds on the LEH blocks limit these errors but do not remove them (production models: lowest $S_1$ of $-\oProdLEHmin{}$\,eV for LEH and $-\oProdSMmin{}$\,eV for Direct-softmin), and the disagreement between seeds identifies them. Systematic errors, such as the too high $S_1$ excitation energies of the larger chromophores, were shared by all seeds of both architectures, so the disagreement between seeds is not a sufficient error estimate far from the training data.

Our results outline the steps toward a universal excited-state MLIP, comparable to the foundation models for the electronic ground state~\cite{batatia2025foundation,levine2025omol25}. The first step is a readout that is size consistent by construction. A ground-state MLIP predicts one extensive energy, whereas an excited-state MLIP predicts a set of states whose number, order, and character change with the size of the system. The latent Hamiltonian defines what a model should predict for local excitations of separated fragments, the union of their spectra, and a matrix with local blocks can still have eigenvectors that extend over a whole conjugated molecule. This readout transferred better to larger, conjugated, and noninteracting molecules, and fine-tuning a pretrained multistate model did not make up for a readout without this property. Extending the representation to charge-transfer states, states delocalized beyond the cutoff, and couplings between states with a physical meaning is the next step on the model side. The second step is data: The learning curves did not saturate, and power laws fitted to them reach an $S_1$--$S_3$ error of 0.1\,eV for random test molecules at \lcMolMin{} to \lcMolMax{} million training molecules, depending on the model, \lcFoldMin{} to \lcFoldMax{} times our training set. In our fixed-budget comparison, the number of labelled geometries limited the accuracy and not their distribution over molecules. Additional data are therefore most useful at geometries where the model is uncertain or where the dynamics goes, as in the excited-state sampling of this work, together with error estimates that also detect the errors shared by all seeds. Beyond the closed-shell molecules with C, H, N, O, and F and the singlet states of our dataset, a universal model needs other elements, charged and open-shell molecules, triplet states for intersystem crossing, and larger conjugated systems, for which our models had their largest errors. Because the model errors are already comparable to those of TD-DFT, the step beyond TD-DFT accuracy needs higher-level reference data, for example through transfer learning from many TD-DFT labels to fewer high-level labels, as for ground-state MLIPs~\cite{smith2019ccsdt}. We release the dataset, the code, and all trained models of this work as building blocks for these steps.

\section*{Methods}
\label{sec:methods}
\subsection*{Local excitation Hamiltonian}
\label{sec:theory}
For a molecule with atomic numbers $\{Z_i\}$ and positions $\mathbf{R}=\{\mathbf{r}_i\}$ we attach $k_i$ latent channels to every atom $i$ ($k_i=1$ in all models of this work), giving a matrix of size $M=\sum_i k_i$.
A shared MACE encoder~\cite{batatia2022mace} produces invariant node features $\mathbf{h}_i$, obtained by concatenating the scalar ($l=0$) channels after each interaction layer. The internal $l=1$ features enter the many-body messages.
The model assembles a real symmetric matrix $\mathbf{H}_\theta(\mathbf{R},Z)\in\mathbb{R}^{M\times M}$ from local blocks,
\begin{align}
  \mathbf{H}_{ii} &= \varepsilon_0\,\mathbf{1}_{k_i} + \operatorname{sym}\!\big[\phi_\text{on}(\mathbf{h}_i)\big], \\
  \mathbf{H}_{ij} &= \tfrac12\big[f_c(r_{ij})\,\phi_\text{pair}(\mathbf{h}_i,\mathbf{h}_j,\mathbf{b}(r_{ij})) + f_c(r_{ij})\,\phi_\text{pair}(\mathbf{h}_j,\mathbf{h}_i,\mathbf{b}(r_{ij}))^{\!\top}\big], \quad i\neq j,
\end{align}
where $\phi_\text{on}$ and $\phi_\text{pair}$ are multilayer perceptrons that output $k\times k$ blocks, $\mathbf{b}$ is a Bessel radial basis, $f_c$ is a smooth polynomial cutoff with radius $r_c^H=5$\,\AA{}, and $\varepsilon_0$ is the mean supervised excitation energy of the training set.
Its ascending eigenvalues $\lambda_1\le\dots\le\lambda_M$ are the vertical singlet excitation energies, and excited-state energies and forces follow as
\begin{equation}
  E_{S_n} = E_{S_0} + \lambda_n, \qquad \mathbf{F}_{S_n} = \mathbf{F}_{S_0} - \nabla_{\mathbf{R}}\lambda_n ,
  \label{eq:forces}
\end{equation}
where $E_{S_0}$ is predicted by an extensive atomic-energy readout on the same encoder.
Away from degeneracies, $\nabla_\mathbf{R}\lambda_n=\mathbf{v}_n^\top(\nabla_\mathbf{R}\mathbf{H})\mathbf{v}_n$ (Hellmann--Feynman), which automatic differentiation of the symmetric eigendecomposition provides exactly.
The blocks are rotation-invariant scalars, so $\mathbf{H}(Q\mathbf{R})=\mathbf{H}(\mathbf{R})$ for any rotation $Q$, and permuting atoms permutes rows and columns of $\mathbf{H}$, leaving the spectrum unchanged.
$\mathbf{H}$ is an effective excitation Hamiltonian. It is not a Kohn--Sham or many-electron Hamiltonian, and its eigenvectors carry no guaranteed diabatic meaning.
Variable-size matrices in a batch are zero-padded to the largest $M$ and the padding rows receive distinct large diagonal values, so they never mix with or fall below physical eigenvalues. A batched eigendecomposition is then exact for each molecule.

\paragraph{Size consistency.}
For two fragments A and B separated by more than the encoder receptive field and $r_c^H$, the features of A do not depend on B, and all pair blocks between A and B vanish. Hence
\begin{equation}
  \mathbf{H}(\text{A}+\text{B}) = \mathbf{H}(\text{A})\oplus\mathbf{H}(\text{B}), \qquad
  \operatorname{spec}\mathbf{H}(\text{A}+\text{B}) = \operatorname{spec}\mathbf{H}(\text{A})\uplus\operatorname{spec}\mathbf{H}(\text{B}),
\end{equation}
where $\uplus$ is the multiset union. This holds exactly and gives $\lambda_1(\text{A}+\text{B})=\min[\lambda_1(\text{A}),\lambda_1(\text{B})]$ and exact degeneracies for identical fragments, while $E_{S_0}$ stays additive. This is size consistency within the sector of local single excitations that the model represents. Charge-transfer states between distant fragments, double excitations, and long-range excitonic couplings are outside this sector.
Pooled multistate readouts do not share this property: sum pooling is extensive, mean pooling averages the fragments, and per-state extremal pooling gives the correct $S_1$ but not the higher roots of the union.

\paragraph{Identifiability and bounded blocks.}
Eigenvalues determine $\mathbf{H}$ only up to orthogonal transformations, so couplings and eigenvectors are not identified. This does not affect adiabatic energies and forces. Only the lowest $m=10$ of the $M$ eigenvalues are supervised, and the sorted assignment of $\lambda_1,\dots,\lambda_m$ to $S_1,\dots,S_m$ keeps unsupervised eigenvalues above $\lambda_m$ at training geometries.
Because nothing bounds the block elements, an out-of-distribution environment can produce a large coupling and drive eigenvalues far below zero. This happened for \nOutlier{} of \nOutlierTotal{} test predictions of the two full-data LEH models (pair couplings of \oPairOutMin{}--\oPairOutMax{}\,eV in the affected geometries, compared with at most \oPairOther{}\,eV in all other test geometries). Such failures can be detected from negative excitation energies and from disagreement between seeds, but they dominate $R^2$ when they occur. The production models therefore use soft bounds, $x\mapsto c\tanh(x/c)$ with $c=12$\,eV for onsite and $c=5$\,eV for pair elements, which leave ordinary blocks practically unchanged. They limit such failures but do not remove them, because many bounded couplings can still add up. The direct readouts also predict negative excitation energies for the affected molecules, which are highly strained cages (Supplementary Note~11).

\subsection*{Encoder, readouts and training}
All models of this work share a MACE encoder~\cite{batatia2022mace} with two interaction layers, 128 channels with $l\le1$ hidden irreducible representations, spherical harmonics up to $l=2$, correlation order 3, eight Bessel functions, and a cutoff of 5\,\AA, and differ only in the excitation readout. Besides LEH, we used three direct multistate readouts on the same encoder. An MLP maps each $\mathbf{h}_i$ to an $m$-vector, which is summed over atoms (Direct-sum, the readout of a conventional multistate MLIP), averaged (Direct-mean), or pooled per state with a learned soft minimum, $\Delta E_n=\sum_i w_{in}e_{in}$ with $w_{in}=\operatorname{softmax}_i(-e_{in}/\tau+s_{in})$ (Direct-softmin, in the spirit of extremal pooling~\cite{malosso2026extremal}). For the direct readouts, the $n$-th lowest output is the $S_n$ excitation energy.
The loss combines the mean squared errors of the ten lowest excitation energies (sorted assignment, weight 1, eV$^2$), of the per-atom $S_0$ energy (weight 100, (eV/atom)$^2$, per-element reference energies from a least-squares fit to the training set), of the $S_0$ forces (weight 10, (eV/\AA)$^2$), and of the $S_1$--$S_3$ forces (weight 10), with excited-state forces from Eq.~\eqref{eq:forces}. Missing labels are masked.
We trained in double precision with AdamW, batch size 16, and a learning rate of $3\times10^{-3}$ with 1000 warm-up updates and cosine decay to 1\,\% over a fixed budget of 100k (up to 10k training molecules), 150k (30k), or 300k updates (full set), and used an exponential moving average of the weights (decay 0.99). We kept the checkpoint with the lowest validation loss. We chose these settings in single-seed LEH runs on 10k and 30k training molecules of the benchmark dataset (batch size, learning rate and schedule, 32 to 128 channels, $L=1$ or 2, and $S_0$ weight 10, 100, or 300), based on the errors and outliers on the test sets of these subsets, and report the $L=2$ encoder as a variant. The direct readouts used the same settings, and X-MACE, MS-ANI, and OMNI-P2x the settings recommended by their authors. Unless stated otherwise, errors of LEH and Direct-softmin are averaged over two (full set) or three seeds.

\paragraph{Production models.}
We trained the production models from scratch on the final dataset including all active-learning geometries (\nTrainGeomProd{} training geometries) with the settings above, 300k updates, soft bounds on the LEH blocks, and a weighted excitation-energy loss. With uniform weights, the production models overestimated $S_1$ on validation geometries with a TD-DFT $S_1$ below 1\,eV by \lgUniformLow{}\,eV on average, and similarly on training geometries. The uniform loss gives little weight to the geometries near $S_1/S_0$ crossings, which make up \pctTrainLowGap{}\,\% of the training set (\pctTrainLowGapTwo{}\,\% below 2\,eV). The weighted loss multiplies the squared error of state $S_n$ of a geometry by $w_n\,[1+4\exp(-\max(\Delta E_1,0.5\,\text{eV})/1\,\text{eV})]$, where $\Delta E_1$ is the TD-DFT $S_1$ excitation energy, $w_n=4,2,2$ for $S_1$--$S_3$, and $w_n=1$ for higher states, and normalises by the sum of the weights. The floor at 0.5\,eV avoids emphasising geometries at which TD-DFT itself becomes unreliable. The weighted loss was motivated by the overestimation of $S_1$ near crossings by the uniform-loss models, and the decision itself used validation data only. We fixed the decision rule before the comparison runs finished and applied it to the mean over both architectures (two seeds each). The mean signed $S_1$ error of validation geometries below 2\,eV had to decrease by at least 30\,\%, and neither the validation $S_1$--$S_3$ error nor the $S_1$ error along the validation trajectories was allowed to increase by more than 5\,\%. The weighted loss met this rule, with changes of $-$\lossAlllowgap{}\,\%, \lossAllval{}\,\%, and \lossAlltraj{}\,\%. For Direct-softmin, all three criteria improved. For LEH, the low-gap error decreased by \lossLEHlowgap{}\,\%, but the error along the validation trajectories increased by \lossLEHtraj{}\,\%, mainly because of one of the two seeds. With four seeds, the weighted loss lowered the mean signed $S_1$ error of the validation geometries below 2\,eV from \pLowGapLEHuni{}\,eV to \pLowGapLEHw{}\,eV for LEH and from \pLowGapSMuni{}\,eV to \pLowGapSMw{}\,eV for Direct-softmin and reduced the errors of both architectures on the active-learning test geometries. For LEH, the errors along the validation and test trajectories and for the ladder molecules were larger on average than with the uniform loss and varied more between seeds (Supplementary Table~10). We kept the weighted loss for both architectures as decided by the rule, because a choice per architecture after these results would have used test data.
We trained four seeds of each production architecture and report the mean $\pm$ standard deviation over seeds and the committee prediction, the mean over the four seeds. The committee was \pCommGainMin{}--\pCommGainMax{}\,\% more accurate than the average seed on the test sets of individual geometries, but gained little along the test trajectories, where the seeds share their errors (Supplementary Table~11). We did not select seeds for reporting. The released default checkpoint of each architecture is the seed with the lowest validation $S_1$--$S_3$ error.

\subsection*{Dataset}
\label{sec:dataset}
\paragraph{Electronic structure.}
We computed all labels with TURBOMOLE~7.5.1~\cite{turbomole2020} at the TD-$\omega$B97X-D/6-31G* level~\cite{wb97xd,pople631gs}, the level of QCDGE~\cite{zhu2024qcdge}, using full linear response (RPA), RI-J, grid m4, SCF convergence $10^{-7}$\,E$_\text{h}$, density convergence $10^{-7}$, and response convergence $10^{-6}$. For each geometry, we computed $E_{S_0}$ and $\mathbf{F}_{S_0}$, the ten lowest singlet excitation energies with oscillator strengths, and analytic gradients of $S_1$, $S_2$, and $S_3$~\cite{furche2002}, each in a separate \texttt{egrad} run with energy-ordered roots. If the Davidson solver did not converge, we repeated the run once with 200 iterations. We masked a force label if its root deviated by more than $2\times10^{-4}$\,E$_\text{h}$ from the ten-root calculation or if the energy in the gradient file was inconsistent with $E_{S_0}+\Delta E_n$. At QCDGE geometries, RPA reproduced the QCDGE excitation energies with a mean absolute deviation of 0.005\,eV (20 molecules), whereas total energies differed by a nearly constant $0.157\pm0.026$\,eV. We therefore reused the QCDGE excitation energies at QCDGE minima but recomputed $E_{S_0}$ and $\mathbf{F}_{S_0}$ there.

\paragraph{Molecules and geometries.}
We drew molecules from all four QCDGE subsets (QM9-, GDB-11-, and PubChemQC-derived, up to ten heavy atoms, closed-shell molecules with C, H, N, O, and F) in a fixed, source-balanced random order, so the dataset grew in tranches without changing earlier entries. Every molecule contributes its QCDGE minimum and up to two thermal geometries from GFN2-xTB~\cite{gfn2} NVT dynamics started at the minimum with a temperature drawn from 300--1000\,K (one frame between 0.5\,ps and 1.5\,ps). We removed geometries whose bonding changed during the dynamics or whose $S_0$ energy lay below that of the minimum of the same molecule (a reaction). In addition, \nTrainExc{} training geometries and the corresponding test geometries come from one round of model-driven excited-state sampling (see below). We used this first version of the dataset, \nGeomTotal{} geometries of \nMolTotal{} molecules, for all benchmark models. Three active-learning rounds (see below) added \nALGeom{} geometries for the production models. Together with the conjugated test sets, the released dataset contains \nGeomAll{} geometries of \nMolAll{} molecules (Supplementary Table~1).

\paragraph{Excited-state sampling.}
Thermal ground-state sampling rarely reaches the twisted or stretched geometries with small $S_1$ excitation energies that excited-state dynamics visits. For 10\,\% of the molecules (hash-selected, all splits) and all larger test molecules, we ran NVE dynamics on $S_1$ (80\,\%) or $S_2$ (20\,\%) of a preliminary LEH model for up to 500\,fs from the minimum with Maxwell--Boltzmann velocities (300--600\,K). A trajectory stopped at a bond change, an $S_1$ excitation energy below 0.5\,eV, or a non-finite energy. We labelled two frames per trajectory with TD-DFT, a random frame and the frame with the largest disagreement between two independently trained models.

\paragraph{Test sets and splits.}
We assigned molecules to training (85\,\%), validation (5\,\%), and test (10\,\%) sets by a hash of the molecule identifier, with all geometries of a molecule in the same set. An additional balanced Bemis--Murcko scaffold hold-out~\cite{bemis1996} removes \nTestScafMol{} molecules of small scaffold families from training (unseen-scaffold test set). Nested training subsets of 1k, 3k, 10k, 30k, and all \nTrainMol{} molecules (\nTrainGeom{} geometries) share one validation subset of \nValMol{} molecules and identical test sets. The larger test molecules are \nLargeMol{} neutral closed-shell molecules with C, H, N, O, and F and 11--20 heavy atoms from the ZINC250k subset~\cite{gomezbombarelli2018} of ZINC~\cite{irwin2005zinc} (50 per heavy-atom count), which we embedded with RDKit~\cite{rdkit} and optimised with GFN2-xTB. Each has its minimum, one thermal geometry, and two $S_1$-sampled geometries. The conjugated test molecules are \nDyes{} dyes from two experimental databases of organic chromophores~\cite{beard2019uvvis,joung2020optical} and \nLadders{} ladder molecules (polyenes, polyacenes, oligophenylenes, and related series) with up to 30 heavy atoms, each at its minimum and one thermal geometry (\nConjGeom{} geometries). We evaluated the sulfur-containing ones separately, because the main models never saw sulfur. For the dye-augmented models, we split the dye scaffolds into training (80\,\%), validation (10\,\%), and test (10\,\%) families. We removed the molecules of all reference trajectories from the training and validation sets of the active-learning rounds and the production models by comparing covalent graphs. The benchmark training set contains three validation molecules (styrene, benzaldehyde, and cinnamaldehyde, at equilibrium and thermal geometries). The training set of the dye-augmented models, from which we fine-tuned the sampling models of rounds 3 and 4, additionally contains azobenzene, HBDI, and DMABN among the training dyes.

\subsection*{Active learning for the production models}
Active learning, in which the models select new geometries for labelling, for example by committee disagreement, is an established way to extend the training data of MLIPs~\cite{westermayr2019chemsci,zhou2025pal}. Three further rounds of model-driven excited-state sampling (rounds 2--4) extended our dataset to the regions that excited-state dynamics reaches. In each round, two sampling models ran 1\,ps of NVE dynamics on $S_1$ (80\,\%) or $S_2$ (20\,\%) of the first model, with Maxwell--Boltzmann velocities at 300--600\,K. The trajectories started from the minima of all molecules with a photoisomerisable motif (an acyclic C=C, C=N, or N=N bond conjugated with an aromatic ring or another double bond, 8,808 molecules), of all larger test molecules, and, from round 3 on, of a new hash-selected 5\,\% of all molecules. A trajectory stopped at a predicted $S_1$ excitation energy below 0.2\,eV, a bond change, or a non-finite energy. We labelled up to two geometries per trajectory with TD-DFT, the one with the largest double-bond twist and the one with the largest disagreement between the two sampling models. For molecules without a motif, we took the most $S_1$-relaxed geometry instead of the most twisted one. Geometries of validation and test molecules formed new test sets, so the earlier test sets remained unchanged. In round 2, we sampled with the two benchmark LEH models. We fine-tuned the sampling models of rounds 3 and 4 for 60k updates (learning rate $10^{-3}$, cosine decay, bounded blocks) on all data up to the previous round, starting from the dye-augmented LEH model and from the sampling models of the previous round, respectively.

After each round, we evaluated models trained on all data up to that round on seven validation molecules that never entered training or validation (styrene, azobenzene, cyclohex-2-enone, (E)-hexatriene, cytosine, benzaldehyde, and cinnamaldehyde). The loop stops when two conditions are met. (i) The $S_1$ dynamics of the models themselves from the validation minima (five trajectories of 1\,ps at 300\,K per molecule) is stable, with finite energies, no bond change, and a total-energy drift below 0.1\,eV. (ii) The mean errors of the $S_1$ excitation energy and of the change of the $S_0$ energy along the TD-DFT $S_1$ trajectories of these molecules are at most 0.2\,eV or improve by less than 10\,\% over the previous round, at the latest after round 4. Production models trained from scratch on all data after round 4 met both conditions, and the loop stopped (Supplementary Note~14).

\subsection*{Baselines}
\label{sec:baselines}
We trained and evaluated all baselines on the same splits, nested training subsets, and test sets as our models. Their predictions are put in adiabatic order, that is, the excited states are sorted by energy and the forces of each state are permuted with its energy. We computed the errors of all models with the same code.

\paragraph{X-MACE.}
We used the implementation of the authors~\cite{barrett2025xmace} (\texttt{AutoencoderExcitedMACE}) with the settings of its documentation: 128 channels with $l\le1$, two interaction layers, correlation order 3, cutoff 5\,\AA, energy and force weights 100, learning rate $10^{-4}$, batch size 10, an exponential moving average of the weights (0.99), single precision, and 150 epochs. In this model, the energies of all states, including $S_0$, are the sorted eigenvalues of a $4\times4$ tridiagonal matrix, which a multilayer perceptron decodes from a 16-dimensional invariant summed over all atoms (a DeepSets autoencoder). The per-atom energy readouts of the encoder are not used. The eigendecomposition creates the cusps of adiabatic surfaces at state crossings, so the network itself only has to fit smooth quantities. Because the decoder acts on a molecule-level sum, the energies are not size extensive. For molecules larger than the training molecules, the predicted $S_0$ forces kept their direction (correlation 0.96--0.97 with the reference) but were systematically too small (least-squares slope 0.77 for 15--19 and 0.66 for 20 heavy atoms, compared with 0.99 for LEH, Supplementary Note~9).
X-MACE masks forces per geometry rather than per state. We therefore trained it on $S_0$--$S_3$, the states with force labels, and it received no forces on the equilibrium geometries (45\,\% of the training set), whose excited-state forces are not part of the dataset. To test whether these label differences explain the comparison, we trained LEH on exactly the labels that X-MACE sees (three supervised states, the same force mask) with 30k training molecules. Its errors were equal to or lower than those of LEH trained on all labels ($S_1$--$S_3$ MAE of 0.177\,eV and 0.182\,eV for random and larger test molecules, compared with 0.205\,eV and 0.184\,eV, and X-MACE 0.199\,eV and 0.246\,eV).

\paragraph{MS-ANI and OMNI-P2x.}
MS-ANI~\cite{martyka2025msani} extends ANI by appending the state index to the atomic environment vectors and sums per-atom energies for every state. OMNI-P2x~\cite{martyka2026omni} uses the same architecture with an additional one-hot input for the level of theory of the training labels and is pretrained on data from several levels of theory. We trained both with MLatom~\cite{dral2024mlatom3} (version 3.25.5) and its defaults, which are identical for the two models: atomic environment vectors with radial and angular cutoffs of 5.2\,\AA{} and 3.5\,\AA, element networks with 160, 128, and 96 neurons and CELU activations, a force weight of 0.1, and a weight of 0.01 on the gaps between adjacent states. When trained from scratch on labels from a single level of theory, the level-of-theory input is constant, so the OMNI-P2x architecture and MS-ANI are the same model up to initialisation and training noise. Their errors were similar for random test molecules ($S_1$--$S_3$ MAE of 0.45\,eV and 0.41\,eV with all training data), and both failed for larger molecules (0.77\,eV and 1.10\,eV, Supplementary Fig.~3). With the full training set, the MLatom training of both models stopped at the 46\,h wall-time limit of our cluster (OMNI-P2x architecture: 393 epochs, compared with 934 and 1,495 epochs with 30k and 10k training molecules), and we used the checkpoint with the lowest validation loss at that point. In the main figures, we show the OMNI-P2x architecture trained from scratch, referred to as OMNI-P2x arch., and the pretrained OMNI-P2x fine-tuned with the protocol of its authors, referred to as OMNI-P2x, fine-tuned.
We compared the pretrained OMNI-P2x model and two fine-tuning protocols in Supplementary Note~3. Without training on our data and with its excitation energies recalibrated linearly on our validation set, the pretrained model reached an $S_1$--$S_3$ MAE of 0.63\,eV for random test molecules, and its $S_1$ forces deviated by 1.2\,eV/\AA. Fine-tuning the three ensemble members with the protocol of the authors (100 epochs, first and third layer of every element network frozen, gap weight 0.01) gave 0.60\,eV with all training data. In this network, the state index enters only the frozen first layer, so the state-dependent offset between the pretraining level of theory and ours cannot adapt. At 30k training molecules, the fine-tuned $S_1$ energies were 0.78\,eV too low on the validation set. With all layers trainable and a gap weight of 1, the offsets disappeared and the error decreased to 0.45\,eV, the accuracy of training from scratch, but the size transferability of the pretrained layers was lost (0.98\,eV for larger molecules, compared with 0.63\,eV with the protocol of the authors). For molecules similar to the training molecules, training from scratch therefore gave the most accurate OMNI-P2x model, and for larger and conjugated molecules, fine-tuning with the protocol of the authors.

\subsection*{Size-consistency test}
We placed sixty pairs of different and sixty pairs of identical equilibrium test molecules at minimum interatomic distances of 2.5--20\,\AA{}, with molecule B randomly rotated and shifted. For each model, we compared the predicted dimer spectrum with the union of the predicted monomer spectra ($S_1$--$S_3$). For ten pairs of each kind, we computed TD-$\omega$B97X-D dimer spectra (twelve roots) at 2.5--15\,\AA{} with the dataset protocol. These spectra contain the physical interaction and serve as the reference for the accuracy of the dimer predictions.

\subsection*{QCDGE benchmark}
As a public benchmark without forces, we trained on QCDGE molecules with up to eight heavy atoms (19,849 molecules, validation 2,205) and tested on 5,000 molecules with nine and 5,000 with ten heavy atoms, including subsets with scaffolds not present in training. A random split of the same training-set size served as the in-distribution control. All models used the settings above with ten supervised excitation energies and no forces.

\subsection*{Excited-state dynamics}
\label{sec:md_methods}
We computed reference adiabatic $S_1$ trajectories with TURBOMOLE at the level of the dataset, using velocity Verlet with a 0.5\,fs step from the GFN2-xTB minimum with Maxwell--Boltzmann velocities at 300\,K, six roots, and the $S_1$ gradient at every step. The trajectories ran for 500--1000\,fs or until TD-DFT broke down near an $S_1/S_0$ crossing, which we detected as an $S_1$ excitation energy below 0.5\,eV or a total-energy jump above 0.1\,eV. The test trajectories, which we did not use for any decision, are acrolein (1\,ps), uracil, trans-stilbene, the green-fluorescent-protein chromophore (Z)-HBDI, and 4-(dimethylamino)benzonitrile (DMABN, 500\,fs each). The validation trajectories, which we used for the active-learning criterion and the choice of the loss only, are styrene, azobenzene, cyclohex-2-enone, (E)-hexatriene, cytosine, benzaldehyde, and cinnamaldehyde. We evaluated the models on every reference frame before TD-DFT breaks down and started model $S_1$ trajectories from the same initial conditions.

\subsection*{Excited-state forces at equilibrium geometries}
The training data contain excited-state forces only at distorted geometries. To test the $S_1$ gradient at the Franck--Condon point, we computed $S_1$--$S_3$ forces at the equilibrium geometries of 2,000 random and all 500 larger test molecules (test only). The recomputed $S_0$ energies agreed with the dataset to 0.002\,meV.

\subsection*{Metrics}
Excitation-energy errors are mean absolute errors (MAEs) over states and geometries in eV, reported for $S_1$--$S_3$ (supervised by all models) unless stated otherwise. In the tables, energy errors are given in meV. Force errors are component-wise MAEs in eV/\AA{} (meV/\AA{} in the tables), and $S_0$ energy errors are per atom. All predictions are in adiabatic order. We used the same evaluation code and test sets for all models, and Supplementary Table~2 lists all values with $R^2$. Uncertainties are standard deviations over seeds. The sampling uncertainty of the test sets is of similar size or smaller. The 95\,\% intervals of the error of a single model from a bootstrap over molecules, with all geometries of a molecule resampled together, are $\pm$0.002--0.01\,eV for the test sets of thermal and $S_1$-sampled geometries. We resampled the ladder molecules by homologous family and the trajectories as a whole, because the errors along a trajectory are correlated over 10--160\,fs.

\subsection*{Use of large language models}
Code, workflow scripts, analyses, and a first draft of the manuscript were written with the assistance of Claude Code by Anthropic under the direction of the author, who checked the results and revised the text.

\section*{Data availability}
The dataset (geometries, $S_0$ energies and forces, ten singlet excitation energies with oscillator strengths, $S_1$--$S_3$ forces, molecule identifiers, splits and all test sets, including all active-learning geometries) and the TURBOMOLE input and output files (without molecular-orbital files) of the \nRawCalc{} calculations made after we began archiving them, including all excited-state-sampled geometries, are available on Zenodo under a CC BY 4.0 licence (\url{https://doi.org/10.5281/zenodo.24042143}). The TD-$\omega$B97X-D $S_1$ reference trajectories of the validation and test molecules are included with their initial conditions and a flag marking the frames after TD-DFT breaks down, together with the TD-DFT spectra of the molecule pairs and the excitation populations of Supplementary Note~7. QCDGE is available from its original publication~\cite{zhu2024qcdge}.

\section*{Code availability}
The LEH-MACE implementation, the checkpoints of all trained models (all seeds, with the seed with the lowest validation $S_1$--$S_3$ error of each production architecture marked as the default), training configurations, analysis and figure scripts are available at \url{https://github.com/aimat-lab/LEH} under the MIT licence and archived on Zenodo (\url{https://doi.org/10.5281/zenodo.24044613}).

\section*{Acknowledgements}
P.F. gratefully acknowledges the computing time made available on the high-performance computer HoreKa 2 at the National High-Performance Computing Center NHR@KIT. This center is jointly supported by the Federal Ministry of Research, Technology and Space and the state governments participating in the National High-Performance Computing (NHR) joint funding program (https://go-nhr.de/our-partners). P.F. acknowledges support by the state of Baden-W\"urttemberg through bwHPC. P.F. acknowledges the Deutsche Forschungsgemeinschaft (DFG, German Research Foundation) under Germany’s Excellence Strategy for funding via the Excellence Cluster “3D Matter Made to Order” (EXC-2082/2-390761711). P.F. acknowledges funding from the Klaus Tschira Stiftung gGmbH (SIMPLAIX 2).

\section*{Author contributions}
P.F. conceived the study, performed the experiments, analyzed the data, and wrote the manuscript.

\section*{Competing interests}
The authors declare no competing interests.

\bibliographystyle{achemso}
\bibliography{refs}
\clearpage
\includepdf[pages=-]{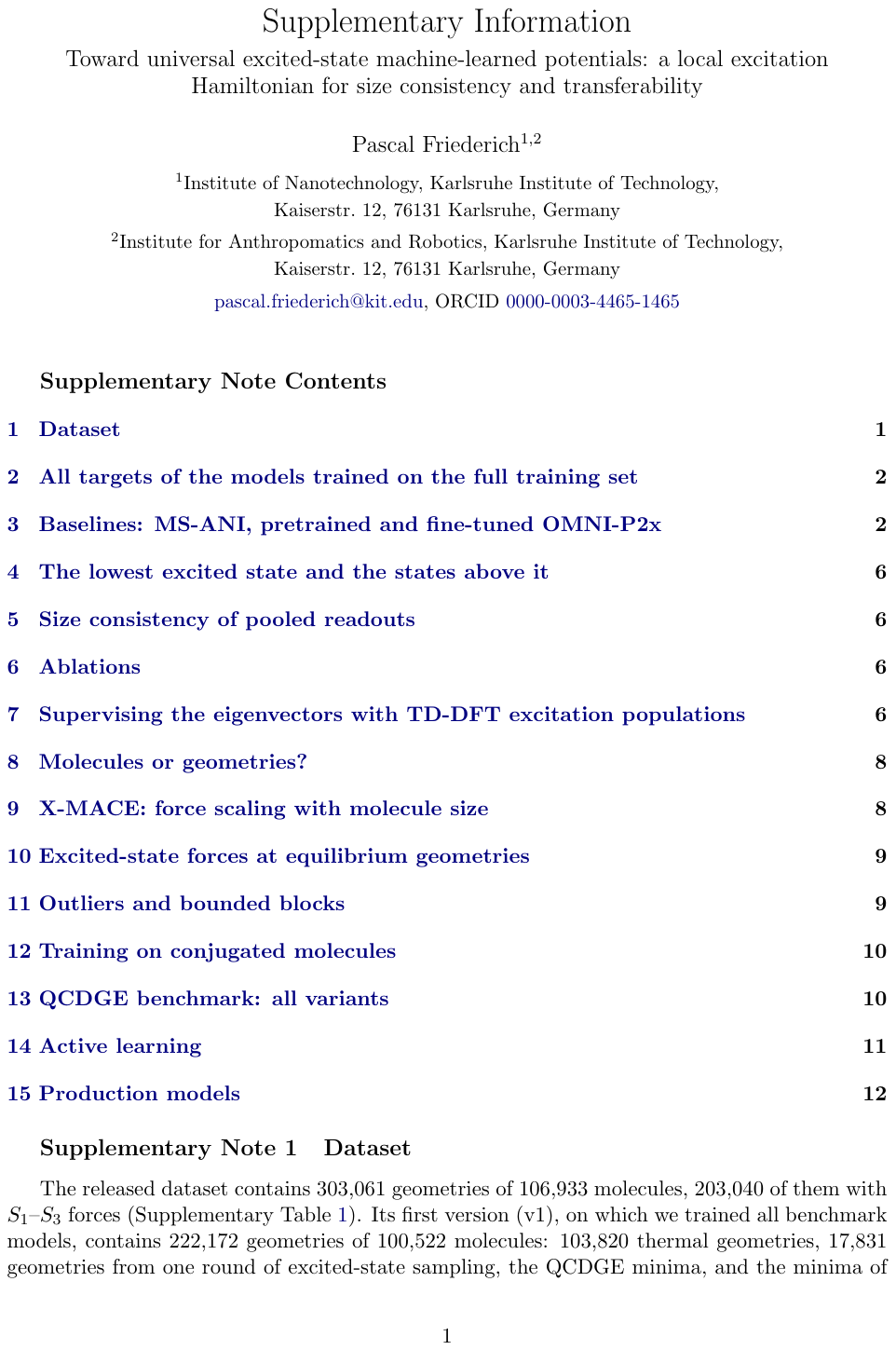}
\end{document}